\documentstyle[twoside]{article}

\newcommand{\tiN}{\raisebox{-6.5pt}{$\displaystyle
\stackrel{\displaystyle N}{\sim}$}}

\newcommand{\beq}{\begin{equation}} \newcommand{\eeq}{\end{equation}}

\catcode`\@=11
\long\def\@makefntext#1{ 
\protect\noindent \hbox to 3.2pt {\hskip-.9pt
$^{{\eightrm\@thefnmark}}$\hfil}#1\hfill} 

\def\thefootnote{\fnsymbol{footnote}}
 \def\@makefnmark{\hbox to 0pt{$^{\@thefnmark}$\hss}}  

\def\ps@myheadings{\let\@mkboth\@gobbletwo
\def\@oddhead{\hbox{} 
\rightmark\hfil\eightrm\thepage}
\def\@oddfoot{}\def\@evenhead{\eightrm\thepage\hfil 
\leftmark\hbox{}}\def\@evenfoot{}
\def\sectionmark##1{}\def\subsectionmark##1{}}

\oddsidemargin=\evensidemargin
\renewcommand{\thefootnote}{\fnsymbol{footnote}}

\newcounter{sectionc}\newcounter{subsectionc}\newcounter{subsubsectionc}
\renewcommand{\section}[1] {\vspace{12pt}\addtocounter{sectionc}{1}
\setcounter{subsectionc}{0}\setcounter{subsubsectionc}{0}\noindent
        {\tenbf\thesectionc. #1}\par\vspace{5pt}}
\renewcommand{\subsection}[1] {\vspace{12pt}\addtocounter{subsectionc}{1}
        \setcounter{subsubsectionc}{0}\noindent
        {\bf\thesectionc.\thesubsectionc. {\kern1pt \bfit #1}}\par\vspace{5pt}}
\renewcommand{\subsubsection}[1] {\vspace{12pt}\addtocounter{subsubsectionc}{1}
        \noindent{\tenrm\thesectionc.\thesubsectionc.\thesubsubsectionc.
        {\kern1pt \tenit #1}}\par\vspace{5pt}}

\newcounter{appendixc}
\newcounter{subappendixc}[appendixc]
\newcounter{subsubappendixc}[subappendixc]
\renewcommand{\thesubappendixc}{\Alph{appendixc}.\arabic{subappendixc}}
\renewcommand{\thesubsubappendixc}
        {\Alph{appendixc}.\arabic{subappendixc}.\arabic{subsubappendixc}}

\renewcommand{\appendix}[1] {\vspace{12pt}
        \refstepcounter{appendixc}
        \setcounter{figure}{0}
        \setcounter{table}{0}
        \setcounter{lemma}{0}
        \setcounter{theorem}{0}
        \setcounter{corollary}{0}
        \setcounter{definition}{0}
        \setcounter{equation}{0}
        \renewcommand{\thefigure}{\Alph{appendixc}.\arabic{figure}}
        \renewcommand{\thetable}{\Alph{appendixc}.\arabic{table}}
        \renewcommand{\theappendixc}{\Alph{appendixc}}
        \renewcommand{\thelemma}{\Alph{appendixc}.\arabic{lemma}}
        \renewcommand{\thetheorem}{\Alph{appendixc}.\arabic{theorem}}
        \renewcommand{\thedefinition}{\Alph{appendixc}.\arabic{definition}}
        \renewcommand{\thecorollary}{\Alph{appendixc}.\arabic{corollary}}
        \renewcommand{\theequation}{\Alph{appendixc}.\arabic{equation}}
        \noindent{\tenbf Appendix \theappendixc #1}\par\vspace{5pt}}
\newcommand{\subappendix}[1] {\vspace{12pt}
        \refstepcounter{subappendixc}
        \noindent{\bf Appendix \thesubappendixc. {\kern1pt \bfit #1}}
        \par\vspace{5pt}}
\newcommand{\subsubappendix}[1] {\vspace{12pt}
        \refstepcounter{subsubappendixc}
        \noindent{\rm Appendix \thesubsubappendixc. {\kern1pt \tenit #1}}
        \par\vspace{5pt}}

\newcommand{\textlineskip}{\baselineskip=13pt}
\newcommand{\smalllineskip}{\baselineskip=10pt}

\def\eightcirc{
\begin{picture}(0,0)
\put(4.4,1.8){\circle{6.5}}
\end{picture}}
\def\eightcopyright{\eightcirc\kern2.7pt\hbox{\eightrm c}}

\newcommand{\copyrightheading}[1]
        {\vspace*{-2.5cm}\smalllineskip{\flushleft
        {\eightrm International Journal of Modern Physics D, #1}\\
        {\eightrm $\eightcopyright$\, World Scientific Publishing
         Company}\\
         }}

\renewenvironment{thebibliography}[1]                   
        {\ninerm
         \baselineskip=11pt                             
         \begin{list}{\arabic{enumi}.}
        {\usecounter{enumi}\setlength{\parsep}{0pt}
         \setlength{\leftmargin 17pt}{\rightmargin 0pt} 
         \setlength{\itemsep}{0pt} \settowidth          
        {\labelwidth}{#1.}\sloppy}}{\end{list}}

\newcounter{itemlistc}
\newcounter{romanlistc}
\newcounter{alphlistc}
\newcounter{arabiclistc}

\newcommand{\fcaption}[1]{
        \refstepcounter{figure}
        \setbox\@tempboxa = \hbox{\eightrm Fig.~\thefigure. #1}
        \ifdim \wd\@tempboxa > 5in
           {\begin{center}
        \parbox{5in}{\eightrm \smalllineskip Fig.~\thefigure. #1 }
            \end{center}}
        \else
             {\begin{center}
             {\eightrm Fig.~\thefigure. #1}
              \end{center}}
        \fi}

\newcommand{\tcaption}[1]{
        \refstepcounter{table}
        \setbox\@tempboxa = \hbox{\eightrm Table~\thetable. #1}
        \ifdim \wd\@tempboxa > 5in
           {\begin{center}
        \parbox{5in}{\eightrm\smalllineskip Table~\thetable. #1 }
            \end{center}}
        \else
             {\begin{center}
             {\eightrm Table~\thetable. #1}
              \end{center}}
        \fi}

\def\@citex[#1]#2{\if@filesw\immediate\write\@auxout    
        {\string\citation{#2}}\fi                       
\def\@citea{}\@cite{\@for\@citeb:=#2\do                 
        {\@citea\def\@citea{,}\@ifundefined             
        {b@\@citeb}{{\bf ?}\@warning
        {Citation `\@citeb' on page \thepage \space undefined}}
        {\csname b@\@citeb\endcsname}}}{#1}}

\newif\if@cghi
\def\cite{\@cghitrue\@ifnextchar [{\@tempswatrue
        \@citex}{\@tempswafalse\@citex[]}}
\def\citelow{\@cghifalse\@ifnextchar [{\@tempswatrue
        \@citex}{\@tempswafalse\@citex[]}}
\def\@cite#1#2{{$\null^{#1}$\if@tempswa\typeout
        {IJCGA warning: optional citation argument
        ignored: `#2'} \fi}}

\def\pmb#1{\setbox0=\hbox{#1}
        \kern-.025em\copy0\kern-\wd0
        \kern.05em\copy0\kern-\wd0
        \kern-.025em\raise.0433em\box0}

\def\fnt#1#2{\footnotetext{\kern-.3em
        {$^{\mbox{\scriptsize #1}}$}{#2}}}

\def\fpage#1{\begingroup
\voffset=.3in
\thispagestyle{empty}\begin{table}[b]\centerline{\footnotesize #1}
        \end{table}\endgroup}

\font\tenbf=cmbx10
\font\tenit=cmti10
\font\tenit=cmti10
\font\bfit=cmbxti10 at 10pt
 1
 1
 1

\font\ninerm=cmr9

\font\eightrm=cmr8

\def\qed{\hbox{${\vcenter{\vbox{                          
   \hrule height 0.4pt\hbox{\vrule width 0.4pt height 6pt
   \kern5pt\vrule width 0.4pt}\hrule height 0.4pt}}}$}}

\begin{document}
\normalsize\textlineskip
{\thispagestyle{empty}
\setcounter{page}{1}

\renewcommand{\thefootnote}{\fnsymbol{footnote}} 

\copyrightheading{Vol. 0, No. 0 (1993) 000--000}

\vspace*{0.88truein}

\fpage{1}
\vspace*{0.035truein}
\centerline{\bf REDUCED PHASE SPACE QUANTIZATION OF}
\centerline{\bf SPHERICALLY SYMMETRIC 
EINSTEIN-MAXWELL THEORY}
\centerline{\bf INCLUDING A COSMOLOGICAL CONSTANT}
\vspace{0.37truein}
\centerline{\footnotesize T. Thiemann\footnote{thiemann@phys.psu.edu}
\footnote{Center for Gravitational Physics
and Geometry, Pennsylvania State University, University Park, PA 16802-6300,
U.S.A.}}
\vspace*{0.015truein}
\centerline{\footnotesize\it Institut f\"ur Theoretische Physik, RWTH Aachen} 
\baselineskip=10pt
\centerline{\footnotesize\it D-52074 Aachen, Germany}
\vglue 10pt

\vspace*{0.21truein}

\begin{abstract}

We present here the canonical treatment of spherically symmetric (quantum)
gravity coupled to  
spherically symmetric Maxwell theory with or without a cosmological constant.
The quantization is based on the reduced phase space which is coordinatized
by the mass and the electric charge as well as their canonically conjugate
momenta, whose geometrical interpretation is explored.\\
The dimension of the reduced phase space depends on the topology chosen,
quite similar to the case of pure (2+1) gravity.\\
We also compare the reduced phase space quantization to the algebraic
quantization. \\
Altogether, we observe that the present model serves as an interesting
testing ground for full (3+1) gravity. \\ 
We use the new canonical variables introduced by Ashtekar 
which simplifies the analysis tremendously.

\end{abstract}

\vspace*{-3pt}\textlineskip

\textheight=7.8truein
\setcounter{footnote}{0}
\renewcommand{\thefootnote}{\alph{footnote}}


The present article summarizes the work of the papers \cite{1} dealing with the
quantization of pure gravity and gravity coupled to a 
spherically symmetric Maxwell field with or without a cosmological constant,
depending on the topology of the the initial data hypersurface.\\
It is self-evident that we can give here only the results, for details the
reader is encouraged to refer to \cite{1}.\\
Throughout we assume that the reader is familiar with the Ashtekar-formulation
of gravity (\cite{2}). Also we use the abstract index formalism and the 
conventions of \cite{3}.\\

To reduce full gravity including matter (\cite{5}) to spherical symmetry,
we require that the 3-metric, the Maxwell electric ($\epsilon^a$) and magnetic 
fields ($\mu^a$)
are Lie annihilated by the generators of the SO(3)
Killing group. That leaves the freedom free that a rotation of the triads with
respect to the tangent bundle is compensated by a rotation with respect to the 
SO(3) bundle. The result of these Killing-reduction prescriptions is the following :\\ 
Denoting the (local) coordinates on the sphere by
$\theta,\phi$ and the radial variable by x as in \cite{1}, we have for the 
gravitational sector ($A_I,\; E^I,\; I=1,2,3$ are angle-independent)
\begin{eqnarray}
(E^x_i,E^\theta_i,E^\phi_i) & = & (E^1 n^x_i \sin(\theta),\frac{\sin(\theta)}
{\sqrt{2}}(E^2 n^\theta_i+E^3 n^\phi_i),\frac{1}{\sqrt{2}}(E^2 n^\phi_i-E^3 
n^\theta_i)),\\
(A_x^i,A_\theta^i,A_\phi^i) & = & (A_1 n^x_i,\frac{1}{\sqrt{2}}(A_2 n^\theta_i
+(A_3-\sqrt{2})n^\phi_i),\frac{\sin(\theta)}{\sqrt{2}}(A_2 n^\phi_i-(A_3-
\sqrt{2})n^\theta_i)\nonumber\\
& & 
\end{eqnarray}
and for the Maxwell sector we obtain ($n^a$ is the standard orthonormal base on
the sphere) 
\beq
(\epsilon^x,\epsilon^\theta,\epsilon^\phi) := (\epsilon(x,t),0,0),\;
(\mu^x,\mu^\theta,\mu^\phi) := (\mu(x,t),0,0) .
\eeq
The Maxwell
potential is thus given by
\beq (\omega_x,\omega_\theta,\omega_\phi)=(\omega(x,t),0,0)
+(\Omega_a(x,t,\theta,\phi)), \eeq
where $\Omega_a$ is a monopole solution with charge $\mu$. 
The cosmological constant will be labelled by the (real) parameter $\lambda$
and by performing a 'duality rotation' we get rid of the magnetic charge. 
\\
The model has thus 4 canonical pairs $(\omega,p\; ;\; A_I,E^I)$ and is
subject to the 4 constraints (so that the reduced phase space is finite
dimensional)
, defined by the following 4 constraint
functionals :
\begin{eqnarray}
^M{\cal G} &=& p'\mbox{ Maxwell-Gauss constraint},\\
^E{\cal G} &=&(E^1)'+A_2 E^3-A_3 E^2\mbox{ Einstein Gauss constraint},\\ 
V &=& B^2 E^3-B^3 E^2\mbox{ Vector constraint},\\ 
C &=&(B^2 E^2+B^3 E^3)E^1+\frac{1}{2}((E^2)^2+(E^3)^2)(B^1
+\kappa\frac{p^2}{2 E^1}+\kappa\lambda E^1)\nonumber\\
& & \mbox{ : Scalar constraint}
\end{eqnarray}
where we have abbreviated the components of the magnetic fields as $B^I$
whose definition is analogous to (0.1) and $E:=(E^2)^2+(E^3)^2$
($\kappa$ is the gravitational constant).\\
As usual, one has to add the ADM energy and the electric charge to the constraint generators in order to make these functionals well-defined.

For spherically symmetric systems, the topology of the 3 manifold is
necessarily of the form $\Sigma^{(3)}=S^2\times\Sigma$ where $\Sigma$ is a
1-dimensional manifold. We will deal here only with asymptotically flat 
topologies.
As is motivated in \cite{1}, we choose 
$ \Sigma=\Sigma_n \; ,\; \Sigma_n\cong K\cup\;\bigcup_{A=1}^n\Sigma_A \; ,$
i.e. the hypersurface is the union of a compact set K (diffeomorphic to a
compact interval) and a collection of ends (each of which is diffeomorphic
to the positive real line without the origin) i.e. asymptotic regions
with outward orientation and all of them are joined to K. 
We want to point out here that the compactum K has {\em nothing} to do
with a horizon, it is just a tool to glue the various ends together and
thus is a kinematical, fixed ingredient of the canonical formalism, whereas
the location of a horizon will depend on the mass of the system which is
a dynamical object. 
Thus, although it is appropriate to
draw the spacetime pictures which one can find in textbooks for, say, the
Schwarzschild configuration with parameter m, the lines $x=m$ which
seperate the 4 Kruskal regions do not, in general, coincide with the (time
evolution of the) compactum K.\\
Boundary conditions for the fields can be derived as in \cite{6} for the 
asymptotic regions. 
In the interiour, the compactum K, we adapt the support of the fields in such
a way that {\em observables} are well-defined. This point is subtle and is 
explained in more detail in \cite{1}.\\

For a review of the method of symplectic reduction the reader is referred
to \cite{7}, for Hamilton-Jacobi methods to \cite{8}..\\
Choosing 'cylinder coordinates'
$ (A_2,A_3)=\sqrt{A}(\cos(\alpha),\sin(\alpha))$,\\$(E^2,E^3)=
\sqrt{E}(\cos(\beta),\sin(\beta))$ 
it is easy to see that the symplectic potential becomes
\beq i\kappa\Theta[\partial_t]=\int_\Sigma dx(\dot{\gamma}\pi_\gamma
+\dot{B^1}\pi_1+\dot{\omega}p(i\kappa)) \; , \eeq
where $\gamma:=A_1+\alpha',\pi_\gamma:=E^1,B^1:=\frac{1}{2}(A-2)\;
\mbox{and}\; \pi_1:=\sqrt{E/A}\cos(\alpha-\beta)$.\\
In the following p will already be taken as a constant. Also we will deal
with an arbitrary cosmological constant for the sake of generality.
We take then the following linear combinations of the vector and the
scalar constraint functional
 $E^1 E^2 V+E^3 C\mbox{ and } 
-E^1 E^3 V+E^2 C $
and set these expressions strongly zero. We then obtain 2 possible
solutions :\\
Case I : $E=0$  corresponds to degenerate metrics and
thus to an unphysical sector.\\
Case II : $E\not =0$ (nondegenerate case)\\
We now conclude
\beq
E^{2/3}=-\frac{2 (E^1)^2}{\kappa(p^2/2+\lambda(E^1)^2+B^1 E^1} B^{2/3}
\eeq and insert this into into the Gauss constraint which can be solved as
follows :
\beq  [\kappa(-p^2+\lambda(E^1)^2/3)+B^1 E^1]^2=m^2 E^1 \; .
\eeq
The integration constant, m, is real and can be shown to coincide with the 
gravitational mass up to a factor.\\
Equation (0.11) is an algebraic equation of fourth order in terms of $E^1$ and,
although algebraically solvable, becomes unpractical to handle in the process of symplectic reduction.
The idea is to change the polarization and to chose $B^1$ as a momentum. Then
(0.11) can be easily solved for $B^1$ and the vector constraint for $\gamma$.
We can then apply the theorem proved in \cite{1} and solve the Hamilton
Jacobi equation by quadrature techniques.
The result for the reduced symplectic potential is given by (modulo a total 
differential)
\beq
(\iota^*\Theta)[\partial_t]=  
 \dot{p}\int_\Sigma dx(-i\frac{p\pi_1}{\pi_\gamma}-\omega)
+\dot{m}\int_\Sigma dx(-i/\kappa)\frac{\pi_1}{\sqrt{\pi_\gamma}}
=: \dot{p}\Phi+\dot{m}T
\eeq
where we have assumed that the cosmological constant is time-independent.\\

The integral expressions for T and $\Phi$ turn out not to be well-defined yet,
one has to add a certain linear combination of constraints in order to achieve
this off the constraint surface. This is satisfactory because a Dirac 
observable is anyway only unique up to a weakly vanishing expression.\\ 
In \cite{1} it is proved that the function $\gamma$ is (weakly) imaginary, hence
$\pi_1$ is imaginary while $\pi_\gamma=E^1$ is real. Accordingly, the
volume parts of T and $\Phi$ are both real.\\ 
The resulting reduced phase space can thus be described as follows : in every
asymptotic end A we have a cotangent bundle over $R^2$.
The situation for K can be handled in a similar manner.\\

Computing Poisson brackets among the observables and between observables
and symmetry generators reaffirms the canonical structure that has been
formally derived above and that the observables are really
gauge invariant when choosing the Lagrange-multipliers of compact support.
For symmetry transformations on the other hand we obtain
\beq
\{m_A,G\}  =  0,\;
\{T_A,G\}  =  N_A,\; 
\{p_A,G\}  =  0,\; 
\{\Phi_A,G\}  =  U_A
\eeq
where we have defined
$ N_A(t):=N(x=\partial\Sigma_A,t)\;,\;U_A(t):=U(x=\partial\Sigma_A,t)$ 
where $N:=\det(q)^{1/2}\tiN$ is the lapse function, $G:=\int dx[\Lambda
(^E{\cal G})+U(^M{\cal G})+N^x V+\tiN C]$ being the constraint 
generator. Hence the observables
are invariant under radial translations and O(2)-rotations at spatial
infinity while they react nontrivially under time-translations and phase
transformations at spatial infinity.\\
For field theories it is 
ot obvious that the equations of motion (0.13) indeed coincide with the
equations of motion that follow from the reduced Hamiltonian ($\bar{\Gamma}$
is the constraint surface)
\beq H_{red}[m,T,p,\Phi]:=G[\Lambda,N^x,M,U]_{|\bar{\Gamma}}=
\sum_{A=1}^n m_A N_A+p_A U_A \;.  \eeq
It is easy to solve the equations of motion (0.13) : introduce functions
$\tau_A(t)\;\mbox{and}\;\phi_A(t)$ defined by
$ \frac{d}{dt}\tau_A=N_A\;\mbox{and}\;\frac{d}{dt}\phi_A=U_A \;.$ 
Then the solution can be written
\begin{eqnarray}
m_A(t)  =  \mbox{const.},\; 
T_A(t)  =  \mbox{const.}+\tau_A(t),\nonumber \\ 
p_A(t)  =  \mbox{const.},\;
\Phi_A(t)  =  \mbox{const.}+\phi_A(t) \; A=1..n
\end{eqnarray}
i.e. the reduced system adopts the form of an integrable system whereby the
role of the action variables is played by the masses and the charges whereas
their conjugate variables take the role of the angle variables.\\
What now is the interpretation of this second set of conjugate variables ?
The interpretation of m and p follows simply from the fact that they can
be derived from the reduced Hamiltonian, i.e. they are the well-known surface
integrals ADM-energy and Maxwell-charge. However, their conjugate partners
are genuine volume integrals and we are not able to write them as known
surface integrals. Nevertheless it is possible to give an interpretation :
In \cite{1} it is shown that T is the eigentime at spatial infinity while
$\Phi$ plays the same role as the variable conjugate to the electric charge
of 1+1 Maxwell theory.

Two objections have occured in the past :\\
1) the variable T vanishes for a Reissner-Nordstrom foliation and by an 
extension of Birkhoff's theorem this foliation can be achieved always. But 
the evolution law (0.13) contradicts Birkhoff's theorem.\\
2) The space of gauge-inequivalent solutions of the Euler-Lagrange
equations is labelled by mass and charge only and thus it is surprising that
the observables $\Phi$ and T exist at all.\\
Both objections can be resolved by the observation that the notions of gauge
are different when one looks at a system either from the Hamiltonian or
Lagrangian viewpoint. This interesting item is analyzed in much more detail in
\cite{1}.\\ 

We finally come to the quantization of the system. We follow the group
theoretical quantization scheme (see ref. \cite{8}). \\
The phase space for every end is just the cotangent bundle over the 
two-plane so the unique Hilbert-space is the usual one : $L_2(R^2,d^2x)$.
The Schroedinger-
equation in the polarization in which
eigentime and the flux act by multiplication and the mass and the charge by
differentiation becomes unambiguously
\beq i\hbar\frac{\partial}{\partial t}\Psi(t;\{T_A\},\{\Phi_A\})
     =(-i\hbar\sum_{A=1}^n[N_A(t)\frac{\partial}{\partial T_A}
     +U_A(t)\frac{\partial}{\partial \Phi_A}])\Psi(t;\{T_A\},\{\Phi_A\}) \; .
\eeq
It can be solved trivially by separation :
$ \Psi(t;\{m_A\},\{\Phi_A\}):=\prod_{A=1}^n\psi_A(t,m_A,\Phi_A)$ 
and by introducing the functions defined by integrating
$ \dot{\tau}_A(t):=N_A(t)\; ,\; \dot{\phi}_A(t):=U_A(t) $.
We then find as the general solution
\beq \psi_A(t,T_A,\phi_A)=C_A\exp(k_A\frac{i}{\hbar}[T_A-\tau_A(t)])
                         \times\exp(l_A\frac{i}{\hbar}[\Phi_A-\phi_A(t)])
\eeq
where $C_A$ is a complex number, whereas $k_A,l_A$ must be
real because the spectrum of the momenta, which are self-adjoint, is real.
These solutions of the time-dependent
Schroedinger equation are obviously peaked at an instant of 'time' t around
the classical solutions (see (0.15)) in the sense that they are strongly 
oscillating off the classical trajectory.

Let us compare with the operator constraint method :\\
We multiply the scalar constraint with a factor of
$E^1$ so that in the ordering in which all momenta (for the Ashtekar
polarization) stand to the right, the scalar constraint becomes
\beq
  [(B^2\frac{\delta}{\delta A_2}+B^3\frac{\delta}{\delta A_3})
\frac{\delta^2}{\delta A_1^2}+\frac{1}{2}(B^1\frac{\delta}{\delta A_1}
+\kappa(-\frac{1}{2}\frac{\delta^2}{\delta\omega^2}
+\lambda\frac{\delta^2}{\delta A_1^2})) 
  (\frac{\delta^2}{\delta A_2^2}+\frac{\delta^2}{\delta A_3^2})]
\Psi[A_I,\omega]=0
\eeq
which is a {\em 4th} order functional differential equation. 
We have here no space to dwell on the issues of ordering and regularization but
it turns out that in the polarization in which the Ashtekar-connection becomes a momentum, the quantization schemes
of Dirac and symplectic reduction actually coincide.\\

Final remark :\\
It is interesting to express the observables found in terms of the 
Ashtekar-connection :
one has simply to solve equation (0.11) for $E^1$
in terms of $B^1$, p and $\lambda$ and plug this into the expressions for
the observables. We restrict for the sake of brevity to the case $\lambda=0$
and obtain
for the integrand of the variables T and
$\Phi+\int_\Sigma dx \omega$ respectively
\beq
-2\frac{A_1+[\arctan(\frac{A_3}{A_2})]'}{(2B^1)^{2-n}}\frac{[p^2\kappa+
\frac{m^2}{B^1}+\sqrt{[p^2\kappa+\frac{m^2}{B^1}]^2-[p^2\kappa]^2})]^{2-n}}
{p^2\kappa+1/2(\frac{m^2}{B^1}+\sqrt{[p^2\kappa+\frac{m^2}{B^1}]^2
-[p^2\kappa]^2})}
\eeq
where $n=1/2$ and $n=1$ respectively and $B^1=1/2((A_2)^2+(A_3)^2-2)$.
This expression is much more complicated than the one in (0.12) in terms
of $\pi_1\;\mbox{and}\;\pi_\gamma$ and it is thus
suggested that in general the polarization that one starts with will not
turn out to be the natural one for the problem at hand.


\begin{thebibliography}{999}


\bibitem{1}  T.\ Thiemann, H.A.\ Kastrup, Nucl.\ Phys.\  {\bf B399}(1993)211 \\
             T.\ Thiemann, H.A.\ Kastrup, PITHA-Preprint 1993 \\
             T.\ Thiemann, PITHA 93-32, August 1993
\bibitem{2}  A.\ Ashtekar,   Phys.\ Rev.\ {\bf D36} (1987)1587; \\
             A.\ Ashtekar,  New Perspepectives in Canonical Gravity
             (Monographs and Textbooks in Physical Science, Bibliopolis,
             Napoli, 1988);\\
             A.\ Ashtekar, Lectures on Non-Perturbative Canonical Gravity
             (World Scientific, Singapore, 1991)
\bibitem{7}  N.\ Woodhouse,  Geometric Quantization
             (Clarendon Press, Oxford, 1980)
\bibitem{4}  C.\ J.\ Isham, A.\ C.\ Kakas, Clas.\ Quantum Grav.\ {\bf 1}
             (1984)621\\
             C.\ J.\ Isham, Class.\ Quantum Grav.\ {\bf 1}(1984)633
\bibitem{6}  R.\ Beig, N.\ o Murchadha, Ann.\ Phys. {\bf 174}(1987)463
\bibitem{8} E.\ T.\ Newman, C.\ Rovelli, Phys.\ Rev.\ Lett.\ {\bf 69}(1992)
             1300
\bibitem{3} R.\ M.\ Wald, General Relativity (The University of Chicago
             Press, Chicago, 1984)
\bibitem{5} A.\ Ashtekar et al, Phys.\ Rev.\ {\bf D40}(1989)2572)

\end{thebibliography}
\end{document}